\documentclass[prl,aps,amssymb,amsmath]{article}
\usepackage{amssymb,amsmath,tikz}
\usepackage{amsthm}
\usepackage{mathrsfs}
\usepackage{pifont}

\showoutput
\newcommand{\ud}{\mathrm{d}}

\theoremstyle{plain}

\newtheorem*{twr*}{THEOREM}
\newtheorem*{lem*}{LEMMA}

\newtheorem*{rem*}{REMARK}

\newtheorem*{notn*}{NOTATION}
\newtheorem*{wiener-ito*}{WIENER-IT\^O-SEGAL DECOMPOSITION}

\begin{document}
\title{{\bf Causal perturbative QED and infra-red states}}
\author{Jaros{\l}aw Wawrzycki \footnote{Electronic address: jaroslaw.wawrzycki@wp.pl or jaroslaw.wawrzycki@ifj.edu.pl}
\\Institute of Nuclear Physics of PAS, ul. Radzikowskiego 152, 
\\31-342 Krak\'ow, Poland}
\maketitle

\vspace{1cm}

\begin{abstract}
At the classical level the electromagnetic field can be
well identified at the spatial infinity. Staruszkiewicz pointed out that the quantization of the electromagnetic field at spatial infinity is essentially unique and follows from the two fundamental principles: 1) gauge invariance and 2) canonical commutation relations for canonically conjugated generalized coordinates, and constructed  a simple and mathematically transparent quantum theory of the Coulomb field, predicting (among other things) a relation between the theory of unitary representations of the $SL(2, \mathbb{C})$ group and the fine structure constant.
Until now this theory has stayed outside the main stream of the perturbative development in QED, mainly due to the unsolved infra-red-type (IR) problems in the perturbative approach.    
Recently however there has been performed a more careful analysis of free fields including the 
mass less free gauge fields, such as the electromagnetic potential field, their Wick
and chronological products, which revealed the need for a more careful and white noise construction of 
these fields, and which opened
a way to resolve IR problems (at least those which shows up at each order separately). Comparison of the perturbatively constructed field at spatial infinity with the quantum phase field of the Staruszkiewicz theory leads to the proof of universality of the unit of charge.

\end{abstract}

\section{Introduction}

Thirty years ago Staruszkiewicz
\cite{Staruszkiewicz1987} (compare also essentially the same text in \cite{Staruszkiewicz1987}) noted that the quantization of the electromagnetic field at spatial infinity is essentially unique and follows from the two fundamental principles: 1) gauge invariance and 2) canonical commutation relations for canonically conjugated generalized coordinates, and constructed  a simple and mathematically transparent quantum theory of the Coulomb field, predicting (among other things) a relation between the theory of unitary representations of the $SL(2, \mathbb{C})$ group and the fine structure constant.
Until now this theory has stayed outside the main stream of the perturbative development in QFT, especially perturbative QED, mainly due to the unsolved infra-red-type problems of the perturbative approach.    
By now the problem of UV-divergences has been solved in physical QFT, such as QED, within the causal perturbative
method.
This method involves the adiabatic switching off the interaction at infinity, as the intermediate stage, and the theory becomes complete, at least at the order-by-order level, when the adiabatic limit of restoring the interaction at infinity is performed. Now in QED we successfully control the adiabatic limit only at the cross section level in each order separately,
as well as for the Green and Wightman functions (free vacuum expectation values) likewise at the order-by-order level. Restoring the interaction
at infinity (in the adiabatic limit) even at each order term separately in the perturbation series for interacting fields
(and not merely for numerical cross sections, or merely for Green or Wightman functions) has until now been unsuccessful.
This is the unsolved infrared (IR) divergence problem. In particular identification of the (interacting) quantum field at spatial infinity, even at the order-by-order level, is an open problem within the causal perturbative QED (not to mention the problem of convergence of the perturbation series). Perhaps the reader well trained in the causal perturbative QED will not even see any necessity
in passing to spatial infinity when dealing with the causal perturbative approach. Let us explain how it arises.

We have solved the infra-red problem in \cite{Wawrzycki2018a} (compare also
\cite{Wawrzycki2018}). 
We have shown there that the lack of success 
in the order-by-order construction of the interacting fields (in the causal perturbative QFT of Bogoliubov, Epstein and Glaser) and of the chronological product of Wick 
polynomials of free fields lies in the careless construction of the free fields especially the Dirac and the 
local mass less gauge fields themselves, such as the electromagnetic potential field within the Gupta-Bleuler (and generally BRST) method, before
passing to their Wick and chronological products. It turns out that their construction with the help of white 
noise Hida operators is necessary (with the clear and explicitly defined action of the $T_4 \circledS SL(2, \mathbb{C})$
in their respective Fock spaces) for the solution of the adiabatic limit problem: construction of the free fields which goes back to Berezin. On the other hand the two most important fields, the Dirac $\boldsymbol{\psi}$ and electromagnetic 
potential $A$ fields, have never been constructed at this level of rigour.  
Is this a real surprise? This is (or would be) not very much surprising, 
at least for some of us, e.g. for Haag who points out on page 48 of his book \cite{Haag} that ``Reviewing our construction of free fields from the irreducible representations of the double covering of the Poincar\'e group 
we notice that the two most important fields,
the Dirac field and the electromagnetic potential, are not directly obtained''. Bogoliubov frequently repeated during his lectures that ``the free electromagnetic field is a true mystery among other free fields''
and some specialists, e.g. Schroer \cite{Schroer},  emphasize that indefiniteness of the inner product (in the Gupta-Bleuler or BRST method) makes the mathematically controllable construction of the free  mass less gauge fields difficult. On the other hand the Gupta-Bleuler or BRST method with indefinite product is necessary for the causal method 
avoiding UV divergences,  so that the problem is real. We have constructed in \cite{Wawrzycki2018} 
the Dirac and the electromagnetic potential fields with the help of white noise Hida operators. 
This allows us to interpret the free fields of the theory as integral kernel operators with vector-valued 
kernels in the sense of Berezin-Obata \cite{obata-book}. But the point is that contributions of all orders 
to interacting fields 
also gain this rigorous interpretation as the free fields themselves: they are well defined 
integral kernel operators in the sense of Berezin-Obata, and moreover they keep this rigorous meaning even if
we replace the ``intensity-of-interaction function'' with a constant $g=1$ equal $1$ everywhere over the space-time.

Although it should be strongly stessed, that the scattering operator, and even the particular higher order contributions to the scattering operator, obtained in our solution by utilizing the Hida operators as the creation-annihilation operators, are not ordinary operators, but generalized operators transforming continously the Hida space into its strong dual. This however is sufficient for the computation of the effective cross section involving many-particle plane wave states of the ``elementary'' fields, which are non-normalizable states but generalized states lying in the space dual to the Hida space. 
In other words we have solved in \cite{Wawrzycki2018a} the so called ``Adiabatic Limit Problem'' in causal
QED, which can be applied to the generalized states, in particular for the high energy scattering involving the many particle plane wave generalized states. But there is also another class of generalized states, which can also be experimentally extracted
and related to the infrared problem, and which can successfully be treated with the causal perturbative QED we have constructed with the help of Hida operators. Namely we consider the generalized (belonging to $(\boldsymbol{E})^*$) homogeneous states of homogeneity degree $-1$ in single particle Fock subspace of the free electromagnetic potential field $A$. Correspondingly to these generalized single particle homogeneous states of the free electromagnetic potential field we have the generalized states in the single particle state spaces of the free massive component fields, which are minimally coupled through the minimal coupling interaction term to the electromagnetic potential field. The structure of the generalized single particle states of the massive free fields, which correspond to the homogeneous single particle states of $A$, depends on the specific type of the massive field, and in particular for the scalar field they are spanned by the states $\varphi$ whose Fourier transform $\widetilde{\varphi}$ has the general form
\[
\widetilde{\varphi}(p)=(p \cdot k)^{-1 + i \nu}, \,\,\, k \cdot k =0, \,\, p \cdot p = m^2, \,\,\,
\nu \in \mathbb{R} \,\,\, \textrm{fixed}.
\] 
The interacting electromagnetic potential field has well defined restriction (let say the homogeneous part of homogeneity $-1$) to the many particle Fock space of the generalized states over the specified class of generalized single particle states (homogeneous of degree $-1$ on the sigle particle subspace of the free electromagnetic potential field) in the total Fock space. This homogeneous of degree $-1$ part of the interacting electromagnetic potential field provides a special realization of the general quatum theory of the electric charge due to \cite{Staruszkiewicz}. There is a unique relationship between the representation structure of the representation of $SL(2, \mathbb{C})$ acting on the specified class of the generalized states, and the value of the fine structure constant, compare Theorem of Subsection 7.6 of \cite{Wawrzycki2018}. This relationship, in particular, and in general the reconstruction of the relationship of causal perturbative QED with \cite{Staruszkiewicz} is completely beyond the scope of the theory which uses the renormalization prescription, involved in handling infinite quantities. The last class of generalized states can be experimentally identified through the Bremsstrahlung phenomena: if we look at the particle which radiates the electromagnetic field due to the acceleration from a suitable distance, at which this process of radiation is practically seen as a scattering at a single point with initial and final four-velocities of the particle equal $u$ and $v$, then the registered radiation will degenrate to the homogeneous solution of Maxwell equations, with the Fourier transform of the correponding electromegnetic potential of the radiation equal
\[
\frac{e}{2\pi} \Big(\frac{u}{u \cdot p} - \frac{v}{v \cdot p} \Big),
\]
and homogeneous of degree $-1$, with $e$ equal to the electric charge of the particle.

In this paper we concetrate on the spectral description (in the sense of Connes) of the gauge group in the space of generalized many particle states, which compose a Fock space of many particle homogeneous
of degree $-1$ states of the field $A$ and the corresponding many particle generalized states of the 
massive fields coupled minimally to $A$. We explain how the very existence of this spectral
construction of the gauge group in the said class of generalized states (composing a Fock-Hilbert space) explains the unversality of the electric charge. Presented proof of universality 
of the charge scale has been suggested long before in \cite{Staruszkiewicz-EssayYA}.

Thus we have obtained causal perturbative QED in which there are no infra-red nor 
ultra-violet infinities, working exclusively with mathematically well defined objects,
which have well defined domains of experimental applicability: A) scattering processes with generalized states of the first class, \emph{i.e.} the many particle plane wave states; and B) the many-particle generalized infra-red states. The method 
of \cite{Wawrzycki2018} can be applied to other realistic QFT, in particular to SM with 
the Higgs field, to which the the Bogoliubov-Epstein-Glaser causal perturbative approach is applicable
(\cite{DKS2}, \cite{DKS3}).      

In particular the first order contribution  
\begin{equation}\label{1-ord-A-g=1}
A_{{}_{\textrm{int}}}^{\mu \,(1)}(g=1,x) = -\frac{e}{4 \pi} \int \ud^3 \boldsymbol{x_{1}} 
\frac{1}{|\boldsymbol{x_1} - \boldsymbol{x}|}
\,
: \overline{\psi} \gamma^\mu \psi : (x_0 - |\boldsymbol{x_1} - \boldsymbol{x}|, \boldsymbol{x_1}).
\end{equation} 
to the interacting electromagnetic field 
$A_{{}_{\textrm{int}}}^{\mu}$ in the adiabatic limit $g=1$ is equal to a finite sum 
\[
A_{{}_{\textrm{int}}}^{\mu \,(1)}(g=1) = \sum \limits_{l,m} \Xi_{l,m}(\kappa_{l,m})
\]
of well defined integral kernel operators 
\[
\Xi_{l,m}(\kappa_{l,m}) \in \mathscr{L}\big( \mathscr{E}, \, \mathscr{L}((\boldsymbol{E}), (\boldsymbol{E})^*) \big)
\]
mapping continuously the space-time test function space $\mathscr{E}$ into the linear topological space 
$\mathscr{L}((\boldsymbol{E}), (\boldsymbol{E})^*)$ of linear continuous operators transforming continuously
the Hida subspace $(\boldsymbol{E})$ (of the tensor product of Fock spaces of $\boldsymbol{\psi}$
and $A$) into its strong dual $(\boldsymbol{E})^*$, equipped with the topology of uniform convergence on 
bounded sets. Moreover some of the integral kernel operator summands  $\Xi_{l,m}(\kappa_{l,m})$
of higher order contributions are much more ``regular'', or even some total higher-order contributions
are more ``regular'' and belong to    
\[
\mathscr{L}\big( \mathscr{E}, \, \mathscr{L}((\boldsymbol{E}), (\boldsymbol{E})) \big).
\]
In particular
\[
A_{{}_{\textrm{int}}}^{\mu \,(1)}(g=1) \in
\mathscr{L}\big( \mathscr{E}, \, \mathscr{L}((\boldsymbol{E}), (\boldsymbol{E})) \big),
\]
compare \cite{Wawrzycki2018} or  \cite{Wawrzycki2018a}.

We remark here that the white noise construction of mass less fields is possible only if the space-time 
test function space $\mathscr{E}$ is equal to the subspace $\mathcal{S}^{00}(\mathbb{R}^4)$ of the Schwartz space 
(no longer equal to the Schwartz space itself 
$\mathcal{S}(\mathbb{R}^4)$ of scalar-, vector-,
spinor-, $\ldots$ functions, depending on the field)  of functions  
$\varphi$ whose Fourier transforms $\widetilde{\varphi}$
belong to $\mathcal{S}^{0}(\mathbb{R}^4)$, \emph{i. e.} to the space $\mathcal{S}^{0}(\mathbb{R}^4)$ of functions 
whose all derivatives vanish at zero. This is remarkable, because in particular for the mass less fields
in Wighman sense the test space can still be chosen to be equal to the ordinary Schwartz space.   

Note that although $\mathcal{S}^{00}(\mathbb{R}^4)$ contains no elements of compact support
it is well suited as a test space for zero mass fields. First note that 
$\mathcal{S}^{00}(\mathbb{R}^4)$ is sufficiently reach to contain element for each arbitrary small 
open conic type-shape set $C$, with the support in $C$. In particular the causal relations may be expressed
within $\mathcal{S}^{00}(\mathbb{R}^4)$. Because the commutator functions of zero mass free fields
are homogeneous, then $\mathcal{S}^{00}(\mathbb{R}^4)$ is also sufficient to provide sufficient
basis for the splitting problem of  these commutator functions and their tensor products,
for the proof compare\footnote{In the following text Subsections will always refer to the reference 
\cite{Wawrzycki2018}.} Subsection 5.7, of \cite{Wawrzycki2018}.

\section{A peculiarity of mass less gauge fields}

For fields which contain zero mass gauge fields (before the interaction is switched on)  
there is one non-trivial problem we are confronted with (one aspect of which shows up already at the 
free field level) not encountered when working with non gauge fields. In case of non gauge fields, when the 
representation $U$ acting in the single particle subspace $\mathcal{H}$, and thus its amplification
$\Gamma(U)$ in $\Gamma(\mathcal{H})$ is unitary, the free field is essentially uniquely, i.e. up to unitary equivalence, determined by its general properties: i.e. by the transformation rule pertinent to the concrete representation $U$
which already includes the ``generalized charges'' pertinent to the field, for example the allowed spin of the 
single particle states, e.t.c.. We should expect of the correctly constructed gauge quantum free fields 
that they are likewise essentially uniquely determined by the corresponding ``generalized charged'' structure
pertinent to the field. But in case of zero mass gauge fields, such as the electromagnetic potential field,
the representation $U$  and its amplification $\Gamma(U)$ is unbounded
and Krein-isometric. The natural equivalence for such representations is the existence of Krein isometric mapping
transforming bi-uniquely and bi-continuously the nuclear space $E$, resp. $(E)$,  into itself, 
and which intertwines the representations. Now this equivalence is weaker 
in comparison to the case of unitary equivalence of non-gauge fields where the continuous
Hilbert space isometry defining the equivalence, and which is continuous on the respective nuclear space,
can be extended to a bounded operator -- in fact even to a unitary operator. This is the problem we are confronted
with already at the free field level. One consequence of this weaker equivalence is the following. 
One can construct two equivalent local electromagnetic potential free fields based on the common nuclear spaces
$E = \mathcal{S}^{0}(\mathbb{R}^3; \mathbb{C}^4)$ and $(E)$ (regarded as functions on the orbit, i.e. on the positive energy cone without the apex, with the spatial components of the momentum as the natural coordinates on the cone without the apex) in the single particle spaces and in the Fock spaces respectively, which have different 
infra-red content. Let us formulate this assertion more precisely. The different representatives of free fields of the same equivalence class are constructed by using different inner products and fundamental symmetry operators on $E$ continuous with respect to nuclear topology on $E$, which after completion with respect to the respective inner products give the respective single particle Hilbert spaces of the respective representatives of the field. In general different representatives of the same equivalence class of the free field may be constructed in this way. The single particle representations
$U$ in case of the two representatives of the free electromagnetic potential field differ substantially; 
in the first case $U$, when restricted to the $SL(2, \mathbb{C})$
subgroup, can be written as a direct integral
(with respect to Hilbert space inner product of the single particle Hilbert space of the corresponding representative of the field)
of representations (in general non unitary) acting naturally on the functions of 
the corresponding homogeneities on the cone, and in the second case of the restriction of $U$ to  $SL(2, \mathbb{C})$ corresponding to the other representative of the free field no such direct integral decomposition is possible. 
For a proof compare \cite{Wawrzycki2018}.
This possibility is no surprise 
as the (unbounded) equivalence operator of representations whose representors of Lorentz hyperbolic rotations
are unbounded does not forces any bounded equivalence for the action of Lorentz representors.
As already noted by Epstein and Glaser, the action of the Lorentz subgroup is of less importance in
causal perturbative approach to QFT (in fact only translational covariance takes a material role
in the perturbative series as well as the spectral behaviour of states of the relevant domains with respect to the 
joint spectrum of translation generators) and on the other hand the (weaker) equivalence for the 
translation generators which are
by construction unitary and Krein-unitary reduces to ordinary unitary and Krein unitary equivalence for the action of the translation representors. Nonetheless sensitivity of the infra-red asymptotic behaviour to the particular choice within one equivalence class of the free field cannot be simply ignored. This is because different asymptotic 
behaviour corresponding to different concrete realizations of the free field within the same equivalence class 
may survive when passing to interacting fields, 
and on the other hand the electromagnetic field has non-trivial infra-red content corresponding to the Coulomb 
interaction, so that its asymptotic behaviour may (and in fact should) reflect important physical properties 
which cannot be ignored. Moreover it cannot \emph{a priori} be excluded (and even it should be expected)
that this asymptotic behaviour is important in fixing the correct choice among different realizations
of the free field within one and the same equivalence class. Therefore in the causal perturbative approach
the condition of Lorentz covariance is not entirely optional when passing to the zero mass gauge field,
such as the interacting electromagnetic potential quantum field, 
which shows up when we treat the field with more care. Moreover a need for the analysis of the (free and interacting)
field at spatial infinity naturally arises.

\section{Interacting field at spatial infinity}

In order to solve this problem we recall that there exists a simple and elegant theory 
of the quantized homogeneous of degree $-1$ part of the electromagnetic potential field $A$, which 
resides at spatial infinity, i.e. at the three dimensional one-sheet hyperboloid, say the three  
dimensional de Sitter space-time, compare \cite{Staruszkiewicz1987} -- \cite{Staruszkiewicz2009}.
At the classical level extraction of the electromagnetic field which resides at spatial infinity is in principle 
unique and well defined, and it is the homogeneous of degree $-1$``part'' of the field $A$
which is free, \cite{GervaisZwanziger}, determined by a scalar $S(x)$ (of ``electric type'') and a scalar $M(x)$
(of ``magnetic type'') on de Sitter 3-hyperboloid fulfilling the homogeneous wave equation on 
de Sitter 3-hyperboloid.   
As shown in \cite{Staruszkiewicz1987} or \cite{Staruszkiewicz} its quantization can be performed within a natural
way with the commutation relations based essentially on the two principles: the gauge invariance and the canonical
commutation relations for the conjugated generalized coordinates, \cite{Staruszkiewicz1987} or 
\cite{Staruszkiewicz}. As shown in \cite{Staruszkiewicz} the phase of the wave function (of the charge carrying
particle, before the second quantization is performed) is the generalized coordinate conjugated to the total
charge, and at the classical level the phase has been determined in \cite{Staruszkiewicz} as equal to the electric part
$S(x)= -e x^\mu A_\mu(x)$ of the field at infinity, with $A_\mu$ homogeneous of degree $-1$ (in general 
distributional solution of d'Alembert equation). The crucial point is that in computing the total
charge we do not need the global solution of the Maxwell equations but need only to know the solution outside the light cone
e.g. knowing the Dirac homogeneous solution of d'Alembert equation (distributional), \cite{Dirac3rdEd} pp.
303-304, which coincides with the ordinary Coulomb potential field outside the light cone is pretty sufficient. In particular the corresponding
field induced on de Sitter 3-hyperboloid by the Dirac homogeneous solution corresponds to the classical Coulomb field
and is determined by the homogeneous of degree $-1$ Coulomb field solution of Maxwell equations at spatial infinity. Therefore 
the standard commutation rules between the phase  and the total charge  (so identified at spatial infinity
with the respective constants in the general scalar solution of the wave equation on de Sitter 3-hyperboloid)
determine uniquely the commutation rules for the scalar field on de Sitter 3-hyperboloid and include the Coulomb field,
\cite{Staruszkiewicz}. In particular it contains the total electric charge as an operator acting in the Hilbert space of the quantum phase field $S(x)$, as a scalar field on de Sitter 3-hyperboloid, and explains discrete character of the charge. This theory is remarkable for several reasons. First it is very simple and mathematically transparent. The paper \cite{Staruszkiewicz} does not enter mathematical analysis of the theory, but the theory of continuous functionals on $\mathcal{S}^{00}(\mathbb{R}^4)$ and $\mathcal{S}^{0}(\mathbb{R}^4)$, respectively over space-time and in momentum space, provide the distributional background for \cite{Staruszkiewicz}, compare Section 7. 
   
In particular the homogeneous Dirac's solution of d'Alembert equation is a well defined distribution over the test space $\mathcal{S}^{00}(\mathbb{R}^4)$ whose Fourier transform is a continuous functional over $\mathcal{S}^{0}(\mathbb{R}^4)$ with the light cone in the momentum space as the support,
for the proof compare Subsection 6.1.  We show in particular that the 
support of the Dirac solution
as a distribution on $\mathcal{S}^{00}(\mathbb{R}^4)$ is equal to that part of space-time which lies 
outside the light cone. Similar property we have for the transversal homogeneous of degree $-1$ 
electric type solutions
of d'Alembert equation generated by the Lorentz transforms of the Dirac solution. 
This solutions extend over to the (here not the correct)
test function space $\mathcal{S}(\mathbb{R}^4)$, but in highly non unique fashion. In general such extensions 
destroy their space-time support which in general cease to be confined to the outside part of the light cone.
When treated as distributions on the correct test space $\mathcal{S}^{00}(\mathbb{R}^4)$ they become uniquely determined with their space-time supports necessary lying outside the light cone, which has very important physical consequences, compare Sect. 6.

We also show (a detailed proof
can be found in Subsection 7.5) that the standard representation of the commutation relations of Staruszkiewicz theory, proposed in \cite{Staruszkiewicz}, can be characterized (among the infinite family of other possible representations) by the condition that in each reference frame the gauge group $U(1)$ can be reconstructed spectrally in the sense of spectral geometry of Connes, by the phase and the charge operators
$V = e^{iS(u)}, D = (1/e) Q$ of his theory, compare Subsection 7.5. For other possible non standard representations of the commutation relations of Staruszkiewicz this would be impossible
with $V= e^{iS(u)}, D = (1/e) \, Q$.  
The standard representation of \cite{Staruszkiewicz} is in fact the one which is actually used in
the subsequent papers \cite{Staruszkiewicz1987}--\cite{Staruszkiewicz2009}. 
Second, it involves the fine structure constant and relates it non trivially
to the theory of irreducible unitary representations of $SL(2, \mathbb{C})$, mainly through the unitary representation
of $SL(2, \mathbb{C})$ acting in the Hilbert space of the quantum phase field $S(x)$, a mathematical theory which have attained a mature form full of computational devices thanks mainly to Gelfand and his school, Neumark, Harish-Chandra and others. In particular (as shown in \cite{Staruszkiewicz1992ERRATUM}) the representation acting in the eigenspace of the total charge operator corresponding to the lowest (regarding the absolute value) non-zero charge contains 
the supplementary series component 
(and if any it must enter discretely) only if the fine structure is sufficiently small. Third, this theory 
contains the quantized Coulomb field (at least as it concerns the asymptotic part outside the light cone).
This is perhaps the most remarkable feature of the theory of Staruszkiewicz \cite{Staruszkiewicz}, 
at least for the perturbative causal approach to QED. Indeed so far as the gauge electromagnetic field was treated with insufficient care the existence of the adiabatic limit was unclear in QED and in particular the status of the Coulomb field
so that the identification of the quantum (interacting) field $A_\textrm{int}(x)$ at spatial infinity was impossible 
within the causal perturbative approach due essentially to the troubles with the adiabatic limit.
But with the electromagnetic potential field treated more carefully we restore the adiabatic limit
and at least in principle we can compute $A_{\textrm{int}}(x)$ as a formal power series in which the switching off
coupling $g(x)$ is moved to infinity, so that the interacting field is now a formal power series in the ordinary 
fine structure constant and not the function $g(x)$, with each order term equal to an operator-valued distribution acting in the Fock space of free fields. This is of capital importance because now we can compare the homogeneous of degree
zero part of the field $x_\mu A_{\textrm{int}}^\mu(x)$ with the quantum phase field $S(x)$ of Staruszkiewicz theory.
For this plan to be realizable we have to learn how to extract a homogeneous part of a fixed homogeneity $\chi$, 
fulfilling d'Alembert equation, of a quantum (interacting) field. Although this task is still non trivial there are several circumstances which both allow the computation to be effective and connect this computation to important physical phenomena. Let us explain this in more details now.
Concerning the extraction of the homogeneous part, fulfilling d'Alembert equation, of a given interacting field, say $x_\mu A_{\textrm{int}}^\mu(x)$, we do it gradually. 

First we observe that a free zero mass field, say a quantum scalar field 
fulfilling d'Alembert equation (or even not necessary fulfilling 
d'Alembert equation, as is the case for $x_\mu A^\mu(x)$, even when $A^\mu(x)$ is free), 
when constructed with the correct test function spaces
$\mathcal{S}^{00}(\mathbb{R}^4)$ and $\mathcal{S}^{0}(\mathbb{R}^4)$ (over space-time and in the momentum picture
respectively), allows a natural construction of a homogeneous part, which is effectively a field on de Sitter
3-hyperboloid, fulfilling d'Alembert equation (or wave equation on de Sitter $3$-hyperboloid which is inhomogeneous in general if the homogeneity degree $\chi \neq 0$). 
Now when looking at the single particle state space we should construct a Hilbert space of homogeneous
(of a fixed degree $\chi$) solutions of d'Alembert equation. In general such solutions have distributional sense and are 
continuous functionals on the test space $\mathcal{S}^{00}(\mathbb{R}^4)$ (with topology inherited from the Schwartz topology on $\mathcal{S}(\mathbb{R}^4)$) and whose Fourier transforms are continuous functionals on the test
space $\mathcal{S}^{0}(\mathbb{R}^4)$ (again with the topology inherited from $\mathcal{S}(\mathbb{R}^4)$) and
have the support concentrated on the (positive sheet) of the cone in the momentum space. Now the restriction to the light cone of the Fourier transforms of these functionals are continuous functionals on the nuclear test space $E
= \mathcal{S}^{0}(\mathbb{R}^3)$ of restrictions of the elements of $\mathcal{S}^{0}(\mathbb{R}^4)$ to the light cone without the apex with the spatial momentum coordinates as the natural coordinates on the cone.
 This gives us a general obstruction 
on the homogeneous generalized single particle states of the homogeneous part of the field we are interested in: 
they should be the continuous functionals 
on the nuclear space $E$, where $E \subset \mathcal{H} \subset E^*$ is the Gelfand triple in the single particle
Hilbert space $\mathcal{H}$ of the initial field in question. Now let us fix a closed subspace $E_{\chi}^{*}$ of 
$E^*$ of functions (functionals) on the cone 
of fixed homogeneity $\chi$. The representation $U$ of the restriction of the double covering of the Poincar\'e group
to the subgroup $SL(2, \mathbb{C})$  acting in $\mathcal{H}$  by the very construction of the field
has the property that each representor maps continuously $E= \mathcal{S}^{0}(\mathbb{R}^3)$ onto $E$ with 
respect to the nuclear topology of $E$
and each representor of its linear dual or transpose (let us denote it by the same sign $U$) transforms 
$E^*$ continuously onto $E^*$ (with its natural strong topology). In particular all elements of $E^*$ of fixed
homogeneity\footnote{In general $\chi$ may assume complex values, although far not all of them are admitted.} 
$\chi$ have a fixed transformation law, let us denote them by $E_{\chi}^{*}$. The representation $U$
 acting on $E_{\chi}^{*}$ is uniquely determined by the action on the homogeneous regular elements of $E_{\chi}^{*}$
i.e functions on the 2-sphere $\mathbb{S}^2$ of unit rays on the cone in momentum space which are smooth on 
$\mathbb{S}^2$. Note that 
$E = \mathcal{S}^{0}(\mathbb{R}^3)$ has the structure of tensor product (for a proof compare Subsect. 
5.6) 
and as a nuclear space is isomorphic
to $\mathcal{S}^{0}(\mathbb{R}) \otimes \mathscr{C}^{\infty}(\mathbb{S}^2)$ and similarly for its dual
$E^* = \mathcal{S}^{0}(\mathbb{R})^* \otimes \mathscr{C}^{\infty}(\mathbb{S}^2)^*$ by the kernel theorem.
 Now using the results of \cite{GelfandV} one can classify all possible Hilbert space inner products
on $E_{\chi}^{*}$ invariant under the representation $U$ of $SL(2, \mathbb{C})$. If such an invariant inner product 
exists for a fixed homogeneity $\chi$ (in general does not exist and if any it is essentially unique) we have to meet our obstruction mentioned to above
before we use it as a single particle inner product of the homogeneous part of the field of homogeneity $\chi$.
Namely it may happen that the closure of $E_{\chi}^{*}$ with respect to this invariant inner 
product leads us out of the space
$E^*$ which is impossible for a field homogeneous of a fixed degree, fulfilling d'Alembert equation, and thus inducing a
field on the de Sitter 3-hyperboloid fulfilling the wave equation on the 3-hyperboloid. If the closure of $E_{\chi}^{*}$
with respect to the invariant inner product lies within the dual space $E^*$, then we obtain a well defined field
when using the closure of $E_{\chi}^{*}$ with respect to the invariant inner product as the single particle Hilbert space by the application of the functor $\Gamma$. The ordinary inverse Fourier transform of the elements of a complete system in this single particle Hilbert space, which are homogeneous (distributional) solutions of d'Alembert equation, determine by restrictions to de Sitter 3-hyperboloid the fundamental modes (waves) fulfilling the wave equation on de Sitter 3-hyperboloid. By the kernel theorem for nuclear spaces\footnote{We are using essentially two types of linear topological spaces: the nuclear spaces and the Hilbert spaces. When writing $E\otimes E$ with nuclear spaces $E$, we mean the projective (coinciding in this case with the equicontinuous) tensor product, which is thus essentially unique, and 
when writing $\mathcal{H} \otimes \mathcal{H}$
for Hilbert spaces $\mathcal{H}$ we mean the Hilbert space tensor product. Note however that the Hilbert space tensor product, projective tensor product and discontinuous tensor product are all different for Hilbert spaces of infinite dimension.} 
$E^* = \mathcal{S}^{0}(\mathbb{R})^* \otimes \mathscr{C}^{\infty}(\mathbb{S}^2)^*$,  
$(E \otimes E)^* = E^* \otimes E^*$ and the Fock structure of the Hilbert space of the homogeneous part of the field
(when it exists at all) is essentially inherited form the Fock structure of the Hilbert space of the initial field itself, and in particular the creation and annihilation operators $a(\widetilde{\varphi})^{+}, a(\widetilde{\varphi})$ of the homogeneous part of the field are well defined, with $\widetilde{\varphi}$ belonging to the closure 
of $E_{\chi}^{*}$ with respect to the invariant Hilbert space inner product, by assumption contained in $E^*$. 
It frequently happens that the spherical harmonics (scalar, spinor, e.t.c, depending on the field)
on the unit 2-sphere of rays on the cone in momentum space, extended by homogeneity, and regarded as elements of 
$E_{\chi}^{*}$ are sufficient to provide a complete system in the single particle subspace 
of the homogeneous part of the field.    

Note in particular that the homogeneous part (fulfilling d'Alembert equation) of a free 
field of degree $\chi$ makes sense only for some particular values
of $\chi$, which of course was to be expected.   

In the next step we observe that the extraction of a homogeneous part of fixed homogeneity $\chi$, fulfilling d'Alembert
equation, of a zero mass field presented above, works also for local free massive fields without any essential changes.
Namely we can extract in a natural way a homogeneous part (of fixed homogeneity $\chi$) fulfilling d'Alembert equation, of a massive local free field. This is possible because the nuclear spaces $\mathcal{S}^{00}(\mathbb{R}^4)$ and 
$\mathcal{S}^{0}(\mathbb{R}^4)$ are closed subspaces of the Schwartz space $\mathcal{S}(\mathbb{R}^4)$
with their topologies inherited from $\mathcal{S}(\mathbb{R}^4)$ (and this holds in any dimension $n$,
i.e. for $\mathcal{S}^{00}(\mathbb{R}^n)$, $\mathcal{S}^{0}(\mathbb{R}^n)$, $\mathcal{S}(\mathbb{R}^n)$).
In case of massive fields the role of the space-time test space is played by functions  (in general
scalar valued, vector valued, spinor valued, depending on the field in question) of $\mathcal{S}(\mathbb{R}^4)$,
and the role of the nuclear space $E$ is played by the Fourier transforms of (scalar valued, vector valued, e.t.c.
depending on the field) functions of $\mathcal{S}(\mathbb{R}^4)$, composing likewise the space $\mathcal{S}(\mathbb{R}^4)$,  restricted to the positive energy sheet 
$\mathscr{O}_{{}_{m,0,0,0}}$ of the two-sheeted mass $m$-hyperboloid, i.e. $E =\mathcal{S}(\mathbb{R}^3)$ (with the spatial components of the momentum as the natural coordinates on  $\mathscr{O}_{{}_{m,0,0,0}}$-- the corresponding orbit of the representation pertinent to the field in question, i.e. just the Lobachevsky space). The point is that 
the representation $U$
in the single particle space of the field in question, uniquely determinates the representation acting on the test space
$\mathcal{S}(\mathbb{R}^4)$,
and thus on $\mathcal{S}^{00}(\mathbb{R}^4)$ -- its closed subspace, and \emph{a fortiori} a representation acting 
on the Fourier transform image and its closed subspace $\mathcal{S}^{0}(\mathbb{R}^4)$, as well as on the restrictions
to the cone of the elements of $\mathcal{S}^{0}(\mathbb{R}^4)$ composing $\mathcal{S}^{0}(\mathbb{R}^3)$, regarded as functions on the cone. In  particular
we have uniquely determined the action of the $SL(2, \mathbb{C})$ subgroup on the elements of 
$\mathcal{S}^{0}(\mathbb{R}^3)^*$, in particular homogeneous distributions in $\mathcal{S}^{0}(\mathbb{R}^3)^*$, whose ordinary inverse Fourier transforms are homogeneous solutions of d'Alembert equation. 
Each distribution $S\in \mathcal{S}^{0}(\mathbb{R}^3)^*$ defines a unique distribution $F$ over 
$\mathcal{S}^0(\mathbb{R}^4)$ concentrated on the (positive sheet $\mathscr{O}$ of the) cone, determined by the condition that $F(\widetilde{\varphi}) = S(\widetilde{\varphi}|_{{}_{\mathscr{O}}})$, well defined because the restriction to $\mathscr{O}$ maps continuously $\mathcal{S}^0(\mathbb{R}^4)$
onto $\mathcal{S}^0(\mathbb{R}^3)$.  
The ordinary Fourier transforms of such $F$-s, regarded as functionals in $\mathcal{S}^0(\mathbb{R}^4)^*$, are homogeneous solutions of d'Alembert equation, in general distributional, i.e. belonging to $\mathcal{S}^{00}(\mathbb{R}^4)^*$.
Thus the representation $U$
induces a unique representation on $\mathcal{S}^{0}(\mathbb{R}^3)^*$ and $\mathcal{S}^{00}(\mathbb{R}^4)^*$.
In particular choosing a subspace of $\mathcal{S}^{0}(\mathbb{R}^3)^*$ of fixed homogeneity $\chi$ we can, 
in case of the scalar field, use the classification of invariant inner products of \cite{GelfandV} on homogeneous functions on the cone, and construct
the homogeneous part of the field of fixed homogeneity $\chi$ as shown above.

The crucial point is that the representation $U$ acting in the single particle subspace of the local massive quantum free field in question, determines a unique representation acting in  $\mathcal{S}^{0}(\mathbb{R}^3)^*$ (regarded as a space of functions on the cone), or resp. in 
$\mathcal{S}^{00}(\mathbb{R}^4)^*$. Moreover the elements of $\mathcal{S}^{0}(\mathbb{R}^3)^*$, or 
$\mathcal{S}^{00}(\mathbb{R}^4)^*$ may in a natural manner be regarded as generalized states of the single particle subspace of the massive field in question, i.e. as elements of $E^* = \mathcal{S}(\mathbb{R}^3)^*$. 
It is instructive to look at this circumstance from the point of view of 
harmonic analysis on the Lobachevsky space -- the orbit $\mathscr{O}_{{}_{m,0,0,0}}$ pertinent to 
the representation $U$ defining the field. Namely we can decompose the restriction of $U$ acting in the single particle space to the subgroup $SL(2, \mathbb{C})$. We obtain a direct integral decomposition into irreducible
(this time $U$ is unitary) sub-representations. Each of the irreducible sub-representations is canonically a representation
acting on functions of fixed homogeneity $\chi$ on the cone. In fact each of the irreducible sub-representations act on Hilbert spaces which up to a measure zero set may be regarded as ordinary functions on the unit sphere 
$\mathbb{S}^2$ of rays on the cone, except for the supplementary series representations (if it enters the decomposition
at all, which is rather exceptional), whose representation space as a complete Hilbert space contains elements which cannot be identified with ordinary functions on the cone. But in each case the elements of the irreducible representation are homogeneous distributions over $\mathcal{S}^{0}(\mathbb{R}^3)$ regarded as the space of restrictions of the elements
$\mathcal{S}^0(\mathbb{R}^4)$ to the positive sheet of the cone. In particular for the massive scalar field we
obtain this assertion without much ado using the decomposition of the representation acting on the scalar functions on the 
Lobachevsky space, acting in the ordinary Hilbert space of square integrable functions with respect to the invariant measure given in \cite{GelfandV}, Ch VI.4.. In order to obtain this theorem in full generality we have to prove that all unitary irreducible representations
of $SL(2, \mathbb{C})$, can be realized on (the closure with respect to an invariant inner product of)
homogeneous functions on the cone. For the spherical-type representations this is already known to be true (the case of the supplementary series representations and the spherical-type representations of the principal series 
has been presented in this manner in \cite{Staruszkiewicz1992ERRATUM}-\cite{Staruszkiewicz2009} and in \cite{GelfandV}). In Subsection 7.2 we give a proof that the closure of the space of homogeneous functions of the supplementary series representation under the invariant inner product is contained within the space $\mathcal{S}^{0}(\mathbb{R}^3)^*$.
It can be however proven for all irreducible unitary representations (and the proof for the remaining irreducible representations easily follows from the results of Subsect. 7.1), or even for all completely irreducible, and not necessary unitary, representations of $SL(2, \mathbb{C})$. In particular any unitary representation 
$(l_0 = m/2, l_1 = i\nu)$, $m \in \mathbb{Z}$, $\nu \in \mathbb{R}$,
of Gefalnd-Minlos-Shapiro \cite{Geland-Minlos-Shapiro} (not necessary spherical-type, i.e. with $l_0$ not necessary equal to zero), can be realized on the space of scalar functions on the cone, homogeneous of degree $-1-i\,\nu$, on using in addition in the transformation formula a homogeneous of degree zero phase factor $e^{i\Theta}$ in the transformation formula, raised to the integer or half-integer power $\pm l_0$ depending on the representation $(l_0, l_1= i\nu)$ we want to achieve, where the phase $e^{i\Theta}$ is the one found in 
\cite{wawrzycki-photon}. The inner product is given by the ordinary $L^2(\mathbb{S}^2)$-norm defined for the restrictions of the homogeneous functions to the unit sphere $\mathbb{S}^2$. 
The decomposition of $U$,
restricted to $SL(2, \mathbb{C})$, can be think of as an application of the 
general Gelfand-Neumark Fourier transform
 corresponding to 
the decomposition of $U$. Because the $SL(2, \mathbb{C})$ group is nonabelian, then the Fourier transform
of a function (even scalar valued) on the Lobachevsky space is no longer scalar valued, but the values of the transform are
homogeneous distributions over $\mathcal{S}^0(\mathbb{R}^3)$ regarded as a space of functions on the cone. 
Each such distribution $S$ defines a unique distribution $F$ over $\mathcal{S}^0(\mathbb{R}^4)$ concentrated on the (positive sheet $\mathscr{O}$ of the) cone, by the condition that $F(\widetilde{\varphi}) = S(\widetilde{\varphi}|_{{}_{\mathscr{O}}})$, well defined because the restriction to $\mathscr{O}$ maps continuously $\mathcal{S}^0(\mathbb{R}^4)$
onto $\mathcal{S}^0(\mathbb{R}^3)$.  
The ordinary Fourier transforms of such $F$-s, regarded as functionals in $\mathcal{S}^0(\mathbb{R}^4)^*$, are homogeneous solutions of d'Alembert equation, in general distributional, i.e. belonging to $\mathcal{S}^{00}(\mathbb{R}^4)^*$.
Now because $E = \mathcal{S}(\mathbb{R}^3)$, the single particle Hilbert space $\mathcal{H}$ and the space of distributions
$E^*$ compose a Gelfand triple $E \subset \mathcal{H} \subset E^*$ (or a rigged Hilbert space), then by 
\cite{GelfandIV}, the elements of the Hilbert space $\mathcal{H}_{{}_{\chi}}$ in the decomposition 
\[
\mathcal{H} = \oplus \int \mathcal{H}_{{}_{\chi}} \ud \sigma(\chi)
\]
corresponding to the decomposition of the representation $U$, restricted to $SL(2, \mathbb{C})$, belong to $E^*$ because the Casimir operators transform $E$ continuously onto itself. Moreover by the analytic continuation  of a distribution, 
\cite{GelfandI}, also the other  distributions homogeneous of degree $\chi$ over $\mathcal{S}^0(\mathbb{R}^3)$, with $\chi$ not entering the decomposition belong to $E^* = \mathcal{S}(\mathbb{R}^3)^*$, i.e. to the space of generalized states of the single particle space of the massive field. Although the application of the general Gelfand-Neumark Fourier transform gives 
a general framework working for all fields, it is not in general computationally useful, because we lose any immediate
relation of the transformation formula of the field, to the respective irreducible components $(l_0, l_1)$ entering the decomposition of $U$ (of course restricted to $SL(2,\mathbb{C})$). For example the explicit realization of the irreducible representation $(l_0, l_1)$ through its action on the scalar homogeneous functions on the cone (with the additional phase factor multiplier $e^{i\Theta}$ raised to the appropriate power) is not much helpful because the phase factor $e^{i\Theta}$ in the transformation formula 
depends on the momentum, which means that its inverse-Fourier-transformed image has non-local transformation law and 
an additional work is needed in recovering the local transformation law of the field in question. In particular taking a direct sum of such irreducible representations with the additional multiplier respectively equal 
$e^{i\Theta}$ and $e^{-i\Theta}$,
acting on functions homogeneous of degree $-1$, we obtain the representation acting in the single particle space 
of a homogeneous of degree $-2$ part of the local 
Riemann-Silberstein quantum vector field, but this is far not obvious, compare \cite{wawrzycki-photon}, \cite{bialynicki-2}, \cite{wawrzycki-photon-app}. 

Therefore in practical computations it is much better to choose another way when computing a homogeneous part of a given
local massive quantum free (scalar, spinor, e.t.c.) field. Namely we construct first the free zero mass counterpart of the
(scalar, spinor, e.t.c.) field. There exists a general construction of such local zero mass fields, compare 
\cite{wawrzycki-photon} or the introductory part of Section 2 (and there is quite a long tradition in constructing such fields, compare e.g. \cite{Bollini}
for the zero mass Dirac field). The homogeneous of degree $-2$ commutator functions of such fields are just the 
quasi asymptotic distributions of the corresponding commutator functions of the massive fields, which we encounter in computing the singularity degree when splitting the massive commutator functions 
due to Epstein-Glaser.
It is not obvious if such counterpart zero mass fields exist, but this is indeed the case at least for
fields we are interested in. Then to the single particle representation of the zero mass (scalar, spinor,
e.t.c.) field, acting on the (scalar, spinor, e.t.c., respectively) functions on the cone (in the momentum picture)
we apply the Gelfand-Graev-Vilenkin Fourier transform
(or its immediate generalization on spinor, etc., valued, functions on the cone) in order to recover the representations acting
on (spinor, tensor, e.t.c.) homogeneous functins, entering the decomposition of this representation. 
The point is that
it is much easer to extend the Gelfand-Graev-Vilenkin Fourier theory on spinor, tensor, etc. valued
functions, then to search at random among the direct summads in the general decomposition 
of the representation $U$ (restricted to $SL(2,\mathbb{C})$) acting in the single particle subspace of the massive field, those which recover the correct 
local transformation formula of the homogeneous part of the field. This task however can be reduced
to the results obtained by Gelfand and Neumark on the classification of unitary representations of 
$SL(2, \mathbb{C})$. The  case of the scalar field we have already at hand without any additional computations.

Summing up we have constructed homogeneous of degree $\chi \in \mathbb{C}$ part of a local free field (working for massive as well as for mass less fields, for non gauge fields and for gauge fields) 
which is well defined only for particular values of
$\chi$. Before we extend this extraction on still more general local fields we should stop for a moment at 
the level of free fields. First note that the introduction of the new class of test spaces 
$\mathcal{S}^{0}(\mathbb{R}^n)$ and $\mathcal{S}^{00}(\mathbb{R}^n)$, essential for the construction of
mass less fields is likewise essential as the distributional basis for \cite{Staruszkiewicz},
compare Section 7. Second note that
the same test spaces are essential in extraction of homogeneous parts of the free fields. A more 
rigorous definition and construction of a homogeneous part of a free field the reader will find in 
Subsection 7.3. And finally let us go back 
to the comparison $x_\mu A^\mu (x) = S(x)$, with the homogeneous of degree zero part of the scalar field
$x_\mu A^\mu (x)$ at the free field level. It turns out that at the free field level,
by extracting of the homogeneous part of degree zero, fulfilling d'Alembert equation, of the field $x_\mu A^\mu (x)$, with $A^\mu$ the free potential, we indeed recover the degenerate case of the theory of Staruszkiewicz, with the fine structure constant put equal to zero, and with the Hilbert space which degenerates to the eigenspace of the total  charge operator corresponding to the eigenvalue zero, compare Subsection 7.4
where we provide a detailed construction. 
Moreover this result holds true for any representative of the free electromagnetic
potential.  Of course this is far not obvious if this results hold true in the full interacting theory and if it is sensitive to the choice of the representative of the free potential as the building block of the causal perturbative series.

As the next step we construct the homogeneous part, in general not fulfilling d'Alembert equation, of a 
local field equal to a Wick polynomial of free fields. Let the homogeneity of the part to be extracted
be $\chi$. In fact we can confine attention to Wick monomials. In particular in order to extract the homogeneous part of the field $:\psi(x)^{n_1} A(x)^{n_2} :$ we sum up 
\[
\sum \limits_{n_1\chi_1 + n_2 \chi_2 = \chi}  \, : \psi_{\chi_1}(x)^{n_1} A_{\chi_2}^{n_2}(x): 
\]
over all fields 
\[
: \psi_{\chi_1}(x)^{n_1} A_{\chi_2}^{n_2}(x):
\]
where $\psi_{\chi_1}(x)$ is the homogeneous part of degree $\chi_1$, fulfilling d'Alembert equation,
of the Dirac field $\psi(x)$ and similarly $A_{\chi_2}(x)$ is the homogeneous of degree $\chi_2$ part
of the field $A(x)$, fulfilling d'Alembert equation;
\[
\begin{split}
\textrm{or we put zero for this sum in case when} \\
\textrm{no $\chi_i$ exist such that $n_1\chi_1 + n_2 \chi_2 = \chi$.}
\end{split}
\]   

As the final step we would like to extract a homogeneous part of an interacting field, especially
$-x_\mu A_{{}_{\textrm{int}}}^{\mu}(x)$. On the other hand the interacting field itself is beyond our reach,
because, although we have given a precise meaning to the limit of the perturbative series in 
\cite{Wawrzycki2018}, we have
leaved so far investigation its convergence. In order to pass over this problem we go back to the 
causal perturbatibe series
for the interacting field $A_{{}_{\textrm{int}}}(x)$ after the adiabatic switching on the interaction
at infinity is performed. Then into each order therm of the causal perturbative series we ``insert'',
in place of each free field operator, its homogeneous part with the respective paring functions replaced by the homogeneous of degree $-2$ zero mass counterparts. Here ``insertion'' means that each integration $d^4x_i$
is replaced with integration over de Sitter hyperboloid and with the homogeneous integrand treated as 
operator distribution on de Sitter hyperboloid. 
Then we confine attention to each order term separately. Next we extract the 
homogeneous part of the chronological product of Wick polynomials of free fields, similarly as for the 
Wick product of free fields, just summing over all summands with factors whose homogeneities sum up to 
$\chi$. 

For example  in order to compute the first order correction
to the homogeneous part $\big(A_{{}_{\textrm{int}}}^{\mu}(x) \big)_{{}_{\chi = -1}}$ 
of homogeneity $\chi = -1$ of the interacting field
$A_{{}_{\textrm{int}}}^{\mu}(x)$, in case when the representative of the free potential field is used which leads to the formulas
for the interacting fields which are given in \cite{Scharf}, Ch. 4.9, 
we need to compute the homogeneous of degree $-1$ part of the generalized operator
(\ref{1-ord-A-g=1}). According to our prescription we ``insert'' the homogeneous of degree $-3$ part of the field
$: \overline{\psi} \gamma^\mu \psi : (x)$ into the formula (\ref{1-ord-A-g=1}); where the ``insertion'' means that the integral in (\ref{1-ord-A-g=1}) of the homogeneous integrand is replaced by the integral over the intersection 
of the space like plane $x_0 = const$ with de Sitter 3-hyperboloid and the integrand
is now regarded as the field on de Sitter 3-hyperboloid, which it naturally induces as a homogeneous field in Minkowski space-time. It is important to note that quantum fields on de Sitter 3-hyperboloid space-time may be integrated over Cauchy surfaces and this integration produces well defined (densely defined) operators in their Hilbert spaces.  We give a proof of it using white noise calculus in Sections 7.4 and 7.3. But the same proof can be performed by using the unitary representation 
of  $SL(2, \mathbb{C})$ acting in the Hilbert space of the homogeneous field. This fact seems to be rather known 
for those who have worked with fields on de Sitter space-time \cite{Staruszkiewicz2004} or on the static 
Einstein Universe space-time \cite{SegalZhouPhi4}, \cite{SegalZhouQED} which are similar in this respect.  

Why is this comparison
\begin{equation}\label{xA=S}
\big(x_\mu A_{{}_{\textrm{int}}}^{\mu}(x)\big)_{{}_{\chi = 0}} = S(x),
\end{equation}
where $S(x)$ is the quantum phase field of Staruszkiewicz theory,  so interesting? First of all
in extracting the homogeneous of degree $-1$ part of the interacting field $A_{{}_{\textrm{int}}}(x)$
only the first and zero order  contributions are non zero:
\[
\big(A_{{}_{\textrm{int}}}^{\mu}(x)\big)_{{}_{\chi = -1}} \,\,\,\,\,\,
= \,\,\,\,\,\, \big(A_{{}_{\textrm{free}}}^{\mu}(x)\big)_{{}_{\chi = -1}} \,\,\,\,\,\, + \,\,\,\,\,\, 
\big(A_{{}_{\textrm{int}}}^{\mu \,(1)}(g=1,x)\big)_{{}_{\chi=-1}},
\]
where $\big(A_{{}_{\textrm{int}}}^{\mu \,(1)}(g=1,x)\big)_{{}_{\chi=-1}}$
is the homogeneous of degree $-1$ part of the generalized operator (\ref{1-ord-A-g=1}) defined 
as above. This is of capital importance. The mechanism which cuts out the higher order terms is 
in principle very simple: the allowed homogeneities $\chi$ for the massive fields
coupled to $A_\mu$ are restricted to relatively small set. In particular the allowed homogeneities
$\chi$ for the scalar massive field are equal: $-1 < \chi <0$ or  $\chi = -1 +i\nu$, $\nu \in \mathbb{R}$,
for the proof compare Subsection 7.2, and Remark 4 of Subsection
7.2.
Similar situation we have for other massive fields, e.g. for the Dirac field.  
On the other hand positive homogeneities for the homogeneous parts of the free field $A_\mu$
are not allowed. In fact we have not finished yet the full classification of allowed homogeneities 
in this case (in Subsection 7.3 we have reduced the classification to application of 
the Gelfand-Graev-Vilenkin method for classification 
of invariant bilinear forms on a nuclear space, and we present some partial results in Subsection 7.3). 
But the assumption that positive homogeneities are impossible is physically reasonable.
On the other hand each factor coming from the retarded (resp. advanced) parts of the commutator
functions contributes additional homogeneity $-2$. Because the number of these factors grows 
together with the order, there remains no room for keeping homogeneity $-1$ of each higher order contribution.     

This in fact is what one should expect, by comparison with the scattering at the classical level in the infra-red regime: the scattered charges produce infra-red electromagnetic 
field but the infra-red electromagnetic field does not scatter charges\footnote{That only first order contribution to
the interacting field at spatial infinity should survive also at the quantum field theory level has been foreseen by Schwinger, as prof. Staruszkiewicz has kindly informed me. Schwinger observes that the only charge carrier fields are massive. The infra-red photons carry to small an energy to produce pairs sufficient to create massive
charge carrying particle. On the other hand we should expect the first order contribution to be non-zero. That there 
persist a kind of ``back-reaction'' we should expect by comparison with the ordinary non relativistic charged quantum particle in the infra-red Bremsstrahlung-type infra-red field: a non-zero contribution to the phase shift will arise for each plane wave of the particle which produces non-trivial change of the packet-type wave function of the particle,
compare e.g. \cite{Staruszkiewicz1981}, \cite{HerdegenMemory}. This is reflected by the non-zero first order contribution.}. 

Moreover,
$\big(A_{{}_{\textrm{int}}}^{\mu \,(1)}(g=1,x)\big)_{{}_{\chi=-1}}$ and $\big(A_{{}_{\textrm{free}}}^{\mu}(x)\big)_{{}_{\chi = -1}}$ by construction commute. 
This again is of capital importance and makes (\ref{xA=S}) still more plausible with 
$x_\mu \big(A_{{}_{\textrm{free}}}^{\mu}(x)\big)_{{}_{\chi = -1}}$ corresponding to 
\begin{equation}\label{clmStar}
\sum \limits_{l=1}^{\infty} \sum \limits_{m = -l}^{m= +l}
\{ c_{lm} f_{lm}^{(+)}(\psi, \theta, \phi) + \textrm{h.c.} \}
\end{equation}
and $x_\mu \big(A_{{}_{\textrm{int}}}^{\mu \,(1)}(g=1,x)\big)_{{}_{\chi=-1}}$ corresponding to 
$-e Q \textrm{th} \psi $ in the expansion of the quantum phase operator
\[
S(\psi, \theta, \phi) = S_0 -e Q \textrm{th} \psi + \sum \limits_{l=1}^{\infty} \sum \limits_{m = -l}^{m= +l}
\{ c_{lm} f_{lm}^{(+)}(\psi, \theta, \phi) + \textrm{h.c.} \} 
\]
of the Staruszkiewicz theory (we are using the notation of  \cite{Staruszkiewicz}). The operator $S_0$
in $S(x)$ is that part which cannot be reproduced by 
$x_\mu \big(A_{{}_{\textrm{int}}}^{\mu}(x)\big)_{{}_{\chi = -1}}$, which again could have been foreseen
by comparison with the classical theory of infra-red fields.

Now the computation of the homogeneous of degree $-3$ part of the free current field 
(in case of the Dirac field coupled to the potential the free current is equal 
$: \overline{\psi} \gamma^\mu \psi : (x)$) is not entirely trivial, 
even in the simpler scalar QED, because there are in general continuum-many
possible homogeneity degrees $\chi$ to play with. Let us explain this in the simpler case of the scalar QED,
where the spinor field $\psi (x)$ is replaced with a scalar (boson) massive complex field, let us denote it
likewise by  $\psi (x)$. In the scalar QED the field $: \overline{\psi} \gamma^\mu \psi : (x)$ is replaced with 
$: \overline{\psi} \overset{\leftrightarrow}{\partial}{}^\mu \psi : (x)$. According to our definition 
each part $\psi_{{}_{\chi}} (x)$ of homogeneity $\chi$ of the scalar field $\psi (x)$ 
contributes to the homogeneous of degree 
$-3$ part of the field $: \overline{\psi} \overset{\leftrightarrow}{\partial}{}^\mu \psi : (x)$
if $\overline{\chi} + \chi = -2$. In particular each homogeneous of degree $-1 - i\nu$, $\nu \in \mathbb{R}$,
part $\psi_{{}_{\chi = -1 -i\nu}} (x)$ of the scalar field $\psi (x)$, has a contribution. 
Each such homogeneous of degree 
$\chi = -1 -i\nu$ part $\psi_{{}_{\chi = -1 -i\nu}} (x)$ of the scalar field $\psi (x)$ is non-trivial, and is constructed on the unitary irreducible representation $U^{\chi = -1+i\nu} = (l_0 = 0, l_1 = i \, \nu)$
of $SL(2, \mathbb{C})$ acting on the homogeneous of degree $\chi = -1 +i\nu$ scalar functions on the cone 
(in the momentum space) as the single particle subspace $\mathcal{H}_{{}_{\chi}}$ of the field 
$\psi_{{}_{\chi = -1 -i\nu}} (x)$,
and is a spherical-type representation of $SL(2, \mathbb{C})$ of the principal series.

Now when working with a finite set of possible homogeneities, say 
$\chi_1 = -1 - i\, \nu_1, \ldots \chi_n = -1 -i \, \nu_n$, 
we need only to consider the finite sum 
\[
: \overline{\psi}_{{}_{\chi_1}} \overset{\leftrightarrow}{\partial}{}^\mu \psi_{{}_{\chi_1}} : (x)
\,\,\, + \,\,\, \ldots
\,\,\, + \,\,\,
: \overline{\psi}_{{}_{\chi_n}} \overset{\leftrightarrow}{\partial}{}^\mu \psi_{{}_{\chi_n}} : (x)
\]
of $n$ independent homogeneous fields 
acting on the tensor product 
$\Gamma(\mathcal{H}_{{}_{\chi_1}}) \otimes \ldots \otimes  \Gamma(\mathcal{H}_{{}_{\chi_n}})$ of their Fock spaces,
by the known property of the functor $\Gamma$:
\[
\Gamma(\mathcal{H}_{{}_{\chi_1}} \oplus \ldots \oplus \mathcal{H}_{{}_{\chi_n}}) =
\Gamma(\mathcal{H}_{{}_{\chi_1}}) \otimes \ldots \otimes  \Gamma(\mathcal{H}_{{}_{\chi_n}}),
\]
as the homogeneous of degree $-3$ part of the field 
$: \overline{\psi} \overset{\leftrightarrow}{\partial}{}^\mu \psi : (x)$.

But already passing from finite set of possible homogeneities to a denumerable
set $\chi_1 , \chi_2 , \ldots $, there arises a subtle point of generalizing the last theorem
to the following 
\[
\Gamma\big(\bigoplus \limits_{n =1}^{\infty} \mathcal{H}_{{}_{\chi_n}}\big) =
\prod \limits_{n \in \mathbb{N}} \otimes \,\, \Gamma(\mathcal{H}_{{}_{\chi_n}}),
\]
where it seems that the $\mathfrak{C}$-adic infinite direct tensor product,
$\prod \limits_{n \in \mathbb{N}} \otimes$, of von Neumann \cite{neumann-inf-tensor} should work here
(although, so far as the author is aware, no proof has until now been performed). And 
when considering the decomposition
\[
U = \int \limits_{\nu > 0} \oplus \,\, U^{\chi = -1+i\nu} \, \ud \nu
\]
of the restriction $U$ of the double covering of the Poincar\'e
group to the $SL(2, \mathbb{C})$ acting in the single particle subspace $\mathcal{H}$ of the scalar field
$\psi(x)$ into irreducible components $U^{\chi = -1+i\nu}$ acting in $\mathcal{H}_{{}_{\chi = -1 + i\nu}}$,
we encounter the following formula
\[
\Gamma\Big(\int \limits_{\nu > 0} \oplus \,\, \mathcal{H}_{{}_{\chi = -1 + i \,\nu}}  \, \ud \nu \Big) =
\prod \limits_{\nu \in \mathbb{R}} \otimes \,\, \Gamma(\mathcal{H}_{{}_{\chi= -1 + i \,\nu}}),
\]
but this time it is far not obvious that the $\mathfrak{C}$-adic infinite direct tensor product
of von Neumann is sufficient here (we expect rather a new infinite tensor product to be needed here);
with (intentionally) infinite system of continuum-many independent fields of respective homogeneities
$\chi = -1 -i\nu$ acting on the corresponding Fock spaces $\Gamma(\mathcal{H}_{{}_{\chi = -1 +i \, \nu}})$.

We propose not to enter these unsolved problems, and confine attention to just one part 
$\psi_{{}_{\chi_1 = -1 -i\nu_1}} (x)$ of $\psi (x)$ of fixed homogeneity $\chi_1 = -1 -i\nu_1$, and then
investigate invariant subspaces of the field

$: \overline{\psi}_{{}_{\chi_1 = -1 -i\nu_1}} 
\overset{\leftrightarrow}{\partial}{}^\mu \psi_{{}_{\chi_1 = -1 -i\nu_1}} : (x)$
(or resp. $: \overline{\psi}_{{}_{\chi_1}} \gamma^\mu \psi_{{}_{\chi_1}} : (x)$).

On the other hand one can show (compare Subsect. 6.4
and 6.6) that the Hilbert space of the quantum phase $S(x)$ of
Staruszkiewicz theory has the following structure 
\begin{equation}\label{HofS}
\mathcal{H} = \mathscr{H}_0 \otimes \mathcal{H}_0.
\end{equation}
Here $\mathscr{H}_0$ is the closed subspace of the Hilbert space $\mathcal{H}$ spanned by  
\[
e^{im S_0}|0\rangle, m \in \mathbb{Z}.
\]
Note that the direct summand with fixed $m$ spanned by
\[
\Big(e^{mS_0}|0\rangle \Big) \otimes \mathcal{H}_0
\] 
in (\ref{HofS}) is the eigenspace of the total charge operator $Q$ corresponding to the eigenvalue $m$.
The direct summand $\mathbb{C} \otimes \mathcal{H}_0 = \mathcal{H}_0$
is the eigenspace corresponding to the eigenvalue zero of $Q$. The Hilbert space $\mathcal{H}_0$
is equal to the Fock space $\mathcal{H}_0 = \Gamma(\mathcal{H}_{0}^{1})$ over 
the single particle space $\mathcal{H}_{0}^{1}$ of ``infra-red transversal photons'' spanned
by 
\[
c_{lm}^{+} |0\rangle.
\]
The representation of $SL(2, \mathbb{C})$ acts on $\mathcal{H}_{0}^{1}$ through the 
Gelfand-Minlos-Shapiro irreducible
representation $(l_0=1, l_1 = 0)$ of the principal series and through its amplification
$\Gamma(l_0=1, l_1=0)$ on $\mathcal{H}_0 = \Gamma(\mathcal{H}_{0}^{1})$, and trivially
on the factor $\mathbb{C}$ in  (\ref{HofS}), \cite{Staruszkiewicz1995}, 
\cite{wawrzycki-HilbertOfPhase}.
The factorization property (\ref{HofS}) is preserved (compare Subsection 6.6) under the representation $U$ 
of $SL(2, \mathbb{C})$ acting in $\mathcal{H}$:
\begin{multline*}
U \mathcal{H} = \big(U\mathscr{H}_0 U^{-1} \big) \otimes \big(U \mathcal{H}_0 U^{-1}\big) \\
= {\mathscr{H}'}_{0} \otimes \big(\Gamma(l_0=1, l_1=0) \mathcal{H}_0 \Gamma(l_0=1, l_1=0)^{-1}\big)
= {\mathscr{H}'}_{0} \otimes \mathcal{H}_0.
\end{multline*}
But under the action of $U$ only the second factor in (\ref{HofS}) is invariant under $U$
where $U$ acts through $\Gamma(l_0=1, l_1=0)$, as said above.
The first factor in (\ref{HofS}) is transformed under $U$ into another subspace 
${\mathscr{H}'}_{0} \subset \mathcal{H}$ spanned by
\[
U e^{im S_0} U^{-1}|0\rangle, m \in \mathbb{Z}.
\]

Finally to the tensor product factorization (\ref{HofS}) of the Hilbert space of the phase field $S(x)$ 
there correspond the tensor product
factorization ${\mathcal{H}'}_1 \otimes {\mathcal{H}'}_0$ of the Hilbert space of the operator
\[
x_\mu\big(A_{{}_{\textrm{int}}}^{\mu}(x)\big)_{{}_{\chi = -1}} \,\,\,\,\,\,
= \,\,\,\,\,\, x_\mu\big(A_{{}_{\textrm{free}}}^{\mu}(x)\big)_{{}_{\chi = -1}} \,\,\,\,\,\, + \,\,\,\,\,\, 
x_\mu\big(A_{{}_{\textrm{int}}}^{\mu \,(1)}(g=1,x)\big)_{{}_{\chi=-1}},
\]
where ${\mathcal{H}'}_0$ is the Fock Hilbert space of the field 
$x_\mu\big(A_{{}_{\textrm{free}}}^{\mu}(x)\big)_{{}_{\chi = -1}}$
and ${\mathcal{H}'}_1$ is the Hilbert space of the field $x_\mu\big(A_{{}_{\textrm{int}}}^{\mu \,(1)}(g=1,x)\big)_{{}_{\chi=-1}}$, by construction equal to an invariant subspace of the Fock space of a homogeneous of degree $-3$ part
$: \overline{\psi}_{{}_{-1 +i\,\nu_1}} \overset{\leftrightarrow}{\partial}{}^\mu \psi_{{}_{-1+i\,\nu_1}} : (x)$ of the field
$: \overline{\psi} \overset{\leftrightarrow}{\partial}{}^\mu \psi : (x)$.
It follows (Subsect. 7.6) that the operators (\ref{clmStar}) and $-e Q \textrm{th} \psi$ on the one hand factorize with respect to the factorization (\ref{HofS}); and on the other hand the operators  
$x_\mu\big(A_{{}_{\textrm{free}}}^{\mu}(x)\big)_{{}_{\chi = -1}}$ and 
$x_\mu\big(A_{{}_{\textrm{int}}}^{\mu \,(1)}(g=1,x)\big)_{{}_{\chi=-1}}$ factorize with respect to the factorization 
${\mathcal{H}'}_1 \otimes {\mathcal{H}'}_0$. 
The Fock space ${\mathcal{H}'}_0$ of the field $x_\mu \big(A_{{}_{\textrm{free}}}^{\mu}(x)\big)_{{}_{\chi = -1}}$
can be naturally identified with the Hilbert space $\mathcal{H}_0$ 
and its action on this space can be naturally identified with the action of the operator
(\ref{clmStar}) on $\mathcal{H}_0$, for the proof compare
Subsect. 7.4 and 7.6.
Both $x_\mu \big(A_{{}_{\textrm{free}}}^{\mu}(x)\big)_{{}_{\chi = -1}}$
and (\ref{clmStar}) act as the unit operator on the respective first factors in 
${\mathcal{H}'}_1 \otimes {\mathcal{H}'}_0$ and respectively
$\mathscr{H}_0 \otimes \mathcal{H}_0$. Similarly $-e Q \textrm{th} \psi$ and 
$x_\mu\big(A_{{}_{\textrm{int}}}^{\mu \,(1)}(g=1,x)\big)_{{}_{\chi=-1}}$
act as the unit operator on the respective second factor, for the proof compare
Subsect. 7.4 and 7.6. 
Thus indeed the operators  
$x_\mu\big(A_{{}_{\textrm{free}}}^{\mu}(x)\big)_{{}_{\chi = -1}}$ and (\ref{clmStar})
understood as operators in the respective Hilbert spaces ${\mathcal{H}'}_1 \otimes {\mathcal{H}'}_0$
and $\mathscr{H}_0 \otimes \mathcal{H}_0$ can be equated, up to a trivial multiplicity. This in particular means that the equality (equivalence)
of the operators $-e Q \textrm{th} \psi$ and $x_\mu\big(A_{{}_{\textrm{int}}}^{\mu \,(1)}(g=1,x)\big)_{{}_{\chi=-1}}$
in their action on the respective first factors would give us full equality (equivalence)
\[
x_\mu\big(A_{{}_{\textrm{free}}}^{\mu}(x)\big)_{{}_{\chi = -1}} = 
\sum \limits_{l=1}^{\infty} \sum \limits_{m = -l}^{m= +l}
\{ c_{lm} f_{lm}^{(+)}(\psi, \theta, \phi) + \textrm{h.c.} \}
\]
 and
\[
x_\mu\big(A_{{}_{\textrm{int}}}^{\mu \,(1)}(g=1,x)\big)_{{}_{\chi=-1}} = -e Q \textrm{th} \psi.
\]

Perhaps the most important reason for the comparison with Staruszkiewicz theory at spatial infinity
lies in giving the realization of the proof for the universality of the unit of charge, outlined in 
\cite{Staruszkiewicz-EssayYA}.  Namely in completing the construction of the subspace invariant for the operator
$x_\mu \big(A_{{}_{\textrm{int}}}^{\mu \,(1)}(g=1,x)\big)_{{}_{\chi=-1}}$ on which indeed it can be identified with
$-e Q \textrm{th} \psi $ for  the potential coupled with various massive charged fields (say, scalar, spinor, e.t.c.) 
with the coupling compatible with gauge invariance, will identify the coupling constant and the charge with the respective constant of the Staruszkiewicz theory. More precisely:
if the equality of $x_\mu \big(A_{{}_{\textrm{int}}}^{\mu \,(1)}(g=1,x)\big)_{{}_{\chi=-1}}$ to the part
$-e Q \textrm{th} \psi$ of phase $S(x)$
of Staruszkiewicz theory is indeed true, then in the Hilbert space
of the field $x_\mu \big(A_{{}_{\textrm{int}}}^{\mu \,(1)}(g=1,x)\big)_{{}_{\chi=-1}}$ 
there must exist the operator $e^{iS_0}$
which together with the operator $(1/e) \, Q$ provides a spectral realization of the global gauge group
$U(1)$. This follows from the fact that this is the case for Strauszkiewicz theory.   
Various contributions to $-e Q \textrm{th} \psi $ coming from various charge carrying fields coupled to the potential $A$ should give the total charge operator $Q$ which together with the corresponding phase 
provides a spectral construction of the global gauge group, as in the case of the Staruszkiewicz theory,
in which $V= e^{iS(u)}, D = (1/e) \, Q$ (or $V=e^{iS_0}, (1/e) \, Q$) define spectrally the 
gauge $U(1)$ group, compare Subsection 7.5). 
This will give us the universality of the unit of charge 
because the various contributions to the global charge $Q$ coming from the various charge carrying fields
all should have common spectrum $e \mathbb{Z}$. Otherwise the total charge operator could not serve as 
the Dirac operator for the $U(1)$ manifold, as the contributions coming from various charge carrying fields
would destroy the spectrum $e \mathbb{Z}$ needed for the spectral reconstruction of the global gauge $U(1)$
group, compare Section 7.5. Thus the common scale for the electric charge comes from the condition that the infra-red fields of each isolated system (involving various charge carrying fields with the couplings to $A$ preserving gauge invariance) provide a spectral description (in their total Hilbert space of infra-red states)
of the global gauge group $U(1)$ as in case of
Stauszkiewicz theory, compare Subsect. 7.5. This mechanism forcing universality of the scale of the electric charge still works even for non standard representation of the commutation rules of the 
Staruszkiewicz theory. The only difference would be in changing of the spectrum of the total charge
$Q$ from $e\mathbb{Z}$ into $ce\mathbb{Z}$ for some constant $c>1$, and in changing
$V= e^{iS(u)}, D = (1/e) \, Q$ into $V=  e^{icS(u)}, D = (1/e) \, Q$ in the spectral construction
of $U(1)$, compare Subsection 7.5
for the definition of the non standard representation.  

Note that in this proof of universality based on the comparison with Staruszkiewicz theory we do not
need to have the operator $S_0$ as constructed in terms of the homogeneous part of the interacting field.
Its computation within the homogeneous part of the interacting fields is perhaps more tricky 
in comparison to $Q$. We suspect that the non-perturbative construction of the causal phase
(compare \cite{Scharf}, Chap. 2.9) will be helpful here, but we had not enough time to try
this way in computations.  

Unfortunately we have not lead the proof that $x_\mu \big(A^{\mu \,(1)}(g=1,x)\big)_{{}_{\chi=-1}}$
is equal to the part $-e Q \textrm{th} \psi $ of $S(x)$  
to an end in this work.   

Perhaps we should remark that various limit operations involved in our computation were commuted rather freely.
Especially we have computed first the homogeneous of degree $-1$ part 
$\big(A_{{}_{\textrm{int}}}^{\mu}(x)\big)_{{}_{\chi = -1}} $ of the field 
$A_{{}_{\textrm{int}}}^{\mu}(x)$ and then constructed homogeneous of degree zero 
part of the field $x_\mu A_{{}_{\textrm{int}}}^{\mu}(x)$ by putting it equal to 
$x_\mu\big(A_{{}_{\textrm{int}}}^{\mu}(x)\big)_{{}_{\chi = -1}}$. 
We did so only to simplify computations, but one should remember that the causal perturbative series
for interacting fields in general does not depend on the order of the following operations: first compute
$A_{{}_{\textrm{int}}}^{\mu}(x)$ and then multiply by $x_\mu$: $x_\mu A_{{}_{\textrm{int}}}^{\mu}(x)$
or first multiply by $x_\mu$: $x_\mu A^\mu(x)$, and then compute  $\big( x_\mu A^\mu(x)\big)_{{}_{\textrm{int}}}$, because $x_\mu$ is a $c$-number.
This order is also unimportant at the free theory level, and it is irrelevant if we first compute the homogeneous of degree 
$-1$ part of the potential field, and then multiply by $x_\mu$, or first multiply by $x_\mu$ and then compute
the homogeneous of degree zero part.

Moreover in extracting the homogeneous part we work effectively with fields on de Sitter 3-hyperboloid 
space-time. On this space-time quantum fields, including interacting fields 
with naturally defined interactions,
behave much better than in the Minkowski space-time. 
Similar fact has been discovered by mathematicians, mainly Segal, Zhou and Paneitz, \cite{SegalZhouPhi4},
\cite{SegalZhouQED}, for the $: \varphi^4:$ theory and for QED on the static Einstein Universe space-time,
with the help of the harmonic analysis on the Einstein Universe, which the authors worked out extensively in a series of papers: \cite{PaneitzSegalI}-\cite{PaneitzSegalIII}. Nonetheless the mechanisms simplifying matters 
are exactly the same for the quantum fields on de Sitter 3-hyperboloid.
In particular the curvature of de Sitter 3-hyperboloid is crucial here (for example QED on the toral compactification of the Minkowski space-time is still very singular, although the set of modes is discrete).

Suppose the operation of multiplication by $x_\mu$ is performed at the very end of the process of computation after the operation of extraction of the homogeneous of degree $-1$ part of the interacting potential, i.e.
for the homogeneous of degree zero part of the field $x_\mu A_{{}_{\textrm{int}}}^{\mu}(x)$
we put $x_\mu \big(A_{{}_{\textrm{int}}}^{\mu}(x)\big)_{{}_{\chi=-1}}$.
It seems that in this computation the various equivalent realizations of the free potential field, 
which are then used in the construction of the perturbative series of the interacting fields, give the same  
$x_\mu \big(A_{{}_{\textrm{int}}}^{\mu}(x)\big)_{{}_{\chi=-1}}$, because all of them have the same pairings. 

Thus passing to infinity is not merely a way to simplify matters but possibly an indispensable step in constructing full theory.

\vspace*{0.2cm}

The author is indebted for helpful discussions to prof. A. Staruszkiewicz. He also wishes
to express his sincere gratitude to Professors A. Staruszkiewicz and M. Je\.zabek for the warm 
encouragement.



\end{document}